# Non-Hermitian bonding and electronic reconfiguration of $Ba_2ScNbO_6$ and $Ba_2LuNbO_6$

Yaorui Tan, Maolin Bo*

Key Laboratory of Extraordinary Bond Engineering and Advanced Materials Technology (EBEAM) of Chongqing, Yangtze Normal University, Chongqing 408100, China

*Author to whom correspondence should be addressed: bmlwd@yznu.edu.cn (Maolin Bo)

**Abstract:**

Despite the extensive applications of perovskite compounds, the precise nature of non-Hermitian bonding in these materials remains poorly understood. In this study, density functional theory calculations were performed to determine the electronic structures of perovskite compounds. In particular, the bandgaps of $Ba_2ScNbO_6$ and $Ba_2LuNbO_6$ were found to be 2.617 and 2.629 eV, respectively, and the deformation bond energies and non-Hermitian bonding of these compounds were calculated. The relationship between the non-Hermitian zeros of the O-Nb bond of $Ba_2ScNbO_6$ and the non-Hermitian zeros of the Sc-O bond was found to be similar but with varying sizes. Further, in-depth research on the non-Hermitian chemistry verified that precise control of atomic bonding and electron states can be achieved, providing new insights into the study of chemical bonds.

## 1. Introduction

A quantum mechanics system is typically an ideal system that is independent of and does not interact with the environment, thus requiring a Hermitian operator (Hamiltonian). In Hermitian systems, the eigenvalues of physical quantities are real numbers that ensure the unitary evolution of the system over time, whereas in non-Hermitian systems they are imaginary; for example, the eigenvalues of time-dependent Hamiltonians are imaginary. In the 1990s, Bender and Boettcher discovered that non-Hermitian operators with parity – time inversion (PT) symmetry produce completely real energy spectra, observable as real numbers[1]. Subsequently, many researchers have explored non-Hermitian systems, uncovering various novel physical properties and phenomena unique to these systems, such as singularities, PT-symmetric phase transitions, probability oscillations, and singular point dynamics. Singularities are unique energy-spectrum degeneracy structures of non-Hermitian systems, resulting in incomplete Hilbert spaces when their corresponding eigenstates merge. The non-conservation of the Fermi – Dirac distribution and the existence of singular points give rise to the special dynamic behaviors of non-Hermitian systems.

Owing to their unique structural properties, perovskite materials exhibit excellent performance and find extensive application in solar cells, optoelectronic devices, catalysts, storage materials, resistors, dielectric materials, and other fields. Different perovskites exhibit different physical properties; for instance, $BaTiO_3$ displays ferroelectricity[2], $SrRuO_3$ ferromagnetism[3], and $LaFeO_3$ weak ferromagnetism[4]. In this study, we investigated the non-Hermitian bonding and electronic configurations of $Ba_2ScNbO_6$[5] and $Ba_2LuNbO_6$[6]. Using density functional theory (DFT), we calculated their density of states (DOS), band structures, and deformation charge densities. Furthermore, we investigated the dynamic behavior of atomic bonding and electron states at singular points in $Ba_2ScNbO_6$ and $Ba_2LuNbO_6$ non-Hermitian systems. Our findings suggest a potential supercritical boundary at the center of the chemical bonding and indicate that precise control of non-Hermitian zeros and bonding can be achieved, providing new ideas for the study of singularity dynamics.

## 2. Methods

## 2.1 DFT calculations

The electronic properties and structural relaxation of $Ba_2ScNbO_6$ and $Ba_2LuNbO_6$ were simulated using the Cambridge Sequential Total Energy Package (CASTEP), which uses DFT with a plane wave pseudopotential and HSE06 to describe the electron exchange and related potentials[7]. The calculations focused on the energy, structure, bonding, and electronic properties of $Ba_2ScNbO_6$ and $Ba_2LuNbO_6$. The cutoff energies, bandgaps, and *k*-point grids used in this study are listed in **Table 1.** All the structures are fully optimized without any symmetry constraints until the force becomes smaller than 0.01 eV/Å and the energy tolerance decreased below $5.0\times10^{-6}$ eV per atom. A convergence threshold of $1.0\times10^{-6}$ eV per atom was selected for the self-consistent field calculations of $Ba_2ScNbO_6$ and $Ba_2LuNbO_6$.

**Table 1** Cut-off energies, bandgaps, and *k*-points of $Ba_2ScNbO_6$ and $Ba_2LuNbO_6$

| | Cut-off energy | *k*-point | Band gap(HSE06) |
|---|---|---|---|
| $Ba_2ScNbO_6$ | 700.0 eV | 10×10×10 | 2.617 eV |
| $Ba_2LuNbO_6$ | 900.0 eV | 10×10×10 | 2.629 eV |

## 2.2 Binding energy and bond charge (BBC) model

BBC models[8] consist of the binding energy model (BC), bond charge model (BC), and Hamiltonian of the BB model **(see Supporting Information)**.

$$H=\xi(0)\sum_{I}\hat{n}_I-t\sum_{I}\sum_{\rho}C_I^{+}C_{I+\rho}=\left(E_n^a-\gamma^m A_n\right)\sum_{I}\hat{n}_I-t\sum_{I}\sum_{\rho}C_I^{+}C_{I+\rho} \tag{1}$$

The binding energy (BE) shifts of the bulk and surface atoms can be expressed as:

$$\Delta E_v(B)=A_n\sum_{I}\hat{n}_I\ ,\quad \Delta E_v(i)=\gamma^m A_n\sum_{I}\hat{n}_I \tag{2}$$

For hybridization bonding ($m\neq1$), Eq. (2) can be written as:

$$\begin{cases} \gamma = \dfrac{E_V(x)-E_V(0)}{E_V(B)-E_V(0)} \\ \gamma^m = \dfrac{E_V(i)-E_V(0)}{E_V(B)-E_V(0)} = \left(\dfrac{E_V(x)-E_V(0)}{E_V(B)-E_V(0)}\right)^m \approx \left(\dfrac{Z_x d_b}{Z_b d_x}\right)^m = \left(\dfrac{Z_b-\lambda_v^{'}}{Z_b}\right)^m \left(\dfrac{d_x}{d_b}\right)^{-m'} \approx \left(\dfrac{d_x}{d_b}\right)^{-m} \end{cases}$$

In $m' = m\left(1-\dfrac{\ln\dfrac{Z_b-\mu_v}{Z_b}}{\ln\left(\dfrac{d_x}{d_b}\right)}\right)$, $\mu_v$ is very small; therefore, $\dfrac{Z_b-\mu_v}{Z_b} \approx 1$ and $m \approx m'$. For compounds, $m \neq 1$.

The Hamiltonian of a system in the BC model is expressed as:

$$H = \sum_{k\sigma} \frac{\hbar^2 k^2}{2m} a_{k\sigma}^{\dagger} a_{k\sigma} + \frac{e_1^2}{2V} \sum_q{}^{*} \sum_{\vec{k}\sigma} \sum_{\vec{K}'\lambda} \frac{4\pi}{q^2} a_{\vec{k}+\vec{q},\sigma}^{\dagger} a_{\vec{K}'-\vec{q},\lambda}^{\dagger} a_{\vec{K}'\lambda} a_{\vec{k}\sigma}$$

(3)

The electron-interaction terms for density fluctuations, which are primarily caused by electrostatic shielding through electron exchange, are given by:

$$\begin{aligned} \delta V_{bc} &= V_{ee}^{'} - V_{ee} \\ &= \frac{e_1^2}{2V} \sum_{\vec{k}} \sum_{\vec{K}'} \sum_{\vec{q}} \sum_{\sigma\lambda} \frac{4\pi}{q^2+\mu^2} a_{\vec{k}+\vec{q},\sigma}^{\dagger} a_{\vec{K}'-\vec{q},\lambda}^{\dagger} a_{\vec{K}'\lambda} a_{\vec{k}\sigma} - \frac{e_1^2}{2V} \sum_q{}^{*} \sum_{\vec{k}\sigma} \sum_{\vec{K}'\lambda} \frac{4\pi}{q^2} a_{\vec{k}+\vec{q},\sigma}^{\dagger} a_{\vec{K}'-\vec{q},\lambda}^{\dagger} a_{\vec{K}'\lambda} a_{\vec{k}\sigma} \\ &= \frac{1}{4\pi\varepsilon_0} \frac{e^2}{2\left|\vec{r}-\vec{r}'\right|} \int d^3r \int d^3r' \rho(\vec{r}) \rho(\vec{r}') e^{-\mu(\vec{r}-\vec{r}')} - \frac{1}{4\pi\varepsilon_0} \frac{e^2}{2\left|\vec{r}-\vec{r}'\right|} \int d^3r \int d^3r' \rho(\vec{r}) \rho(\vec{r}') \end{aligned}$$

The deformation bond energy $\delta V_{bc}$ can be represented as:

$$\delta V_{bc} = \frac{1}{4\pi\varepsilon_0} \frac{e^2}{2\left|\vec{r}-\vec{r}'\right|} \int d^3r \int d^3r' \delta\rho(\vec{r}) \delta\rho(\vec{r}')$$

(4)

### 2.3 **Riemann sphere**

Consider the Euclidean space $R^3$ with coordinates (X, Y, Z ), where the $XY$ plane represents complex set $C^9$. We denote as $S$ a sphere centered at (0,0,1/2) with a radius of 1/2 (i.e., diameter of one unit) and one of its vertices located at the origin of the complex plane, as shown in **Fig. 1**. $N$ = (0,0,1) denotes the north pole of the sphere.

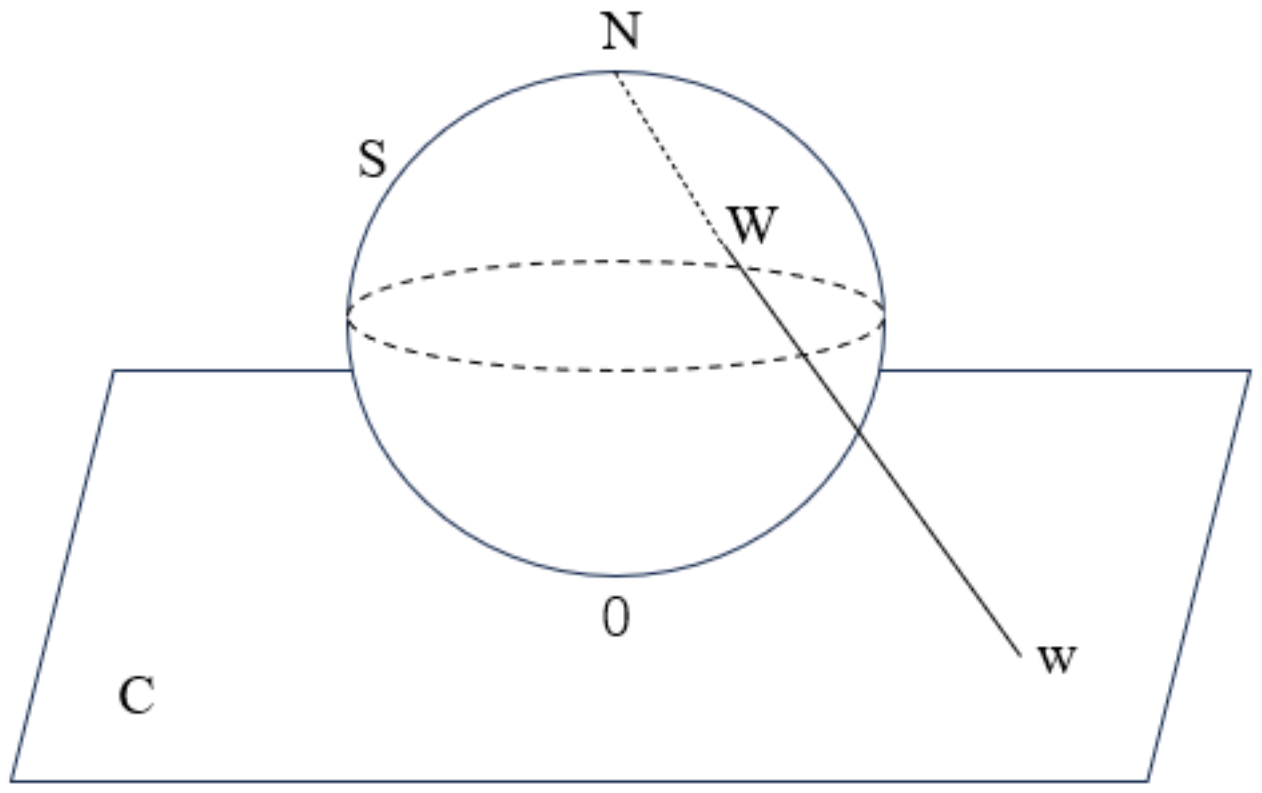

**Fig. 1** Riemann sphere *S* and projection

Assume a point *W* = (X, Y, Z) on sphere *S* other that the north pole. The line $w = x + iy$ connecting points *N* and *W* intersects the *XY* plane at point $w$, which is the projection point of *W* (see **Fig. 1**). Conversely, the $w=(x, y, 0)$ line connecting *N* and any point $w$ on complex set *C*, intersects the sphere at a unique point, denoted as *W*. This geometric explanation provides an objective representation of a point on a perforated spherical surface *S*-{*N*} and a complex plane. The analytical description of this objective representation is as follows.

$$x = \frac{X}{1-Z},\ y = \frac{Y}{1-Z} \tag{5}$$

Substituting *W* with *w*:

$$Y = \frac{y}{x^2+y^2+1}\ X = \frac{x}{x^2+y^2+1},\ Z = \frac{x^2+y^2}{x^2+y^2+1} \tag{6}$$

Intuitively, a complex plane can cover a spherical surface with a hole *S*–{*N*}.

Point *w* approaching infinity on the complex plane corresponds to point *W* infinitely approaching *N* on the sphere; therefore, the north pole, *N*, is the so-called "infinite point". If *N* is defined as an infinite point on sphere *S*, the extended complex plane can be visualized as the entire two-dimensional sphere *S*, known as the Riemann sphere. This explanation enables the mapping of an unbounded set *C* onto a compact set *S* by adding a point; hence, the Riemannian sphere is sometimes referred to as the single-point compactification of a complex set *C*.

Consequently, although infinite points on the sphere can be separated from the complex set, all points on the complex set correspond uniquely with points on the sphere other than the elemental infinite points. In particular, a meromorphic function defined on the extended complex plane can be considered as a mapping of sphere *S* onto itself, transforming the image of the poles to an easy-to-handle point on *S*, i.e., north pole *N*. Therefore, the Riemann sphere provides a robust geometric framework for the composition of complex set *C* and the determination of meromorphic functions.

**2.4 Mobius transformation**

For a point *z* on a complex plane, the Mobius transformation[10] can be expressed as $M(z)=\frac{az+b}{cz+d}$, where *a*, *b*, *c*, and *d* are complex numbers and $ad-bc\neq 0$. This transformation can be regarded as a fractional linear function that maps each point on a complex plane to a unique point.

Each Mobius transformation *M*(*z*) corresponds to a $(2\times 2)$ matrix, [*M*], with complex elements:

$$M(z)=\frac{az+b}{cz+d}\rightarrow[M]=\begin{bmatrix} a & b \\ c & d \end{bmatrix}$$

(7)

Note that complex numbers *z=x+iy* are written as $[\kappa_1,\kappa_2]$, where the ratio $z=\frac{\kappa_1}{\kappa_2}$ of the ordered complex numbers is symmetric to the $z$ secondary coordinates of the complex numbers. As $R^2$ is used to represent (*x*, *y*) sets of real pairs, $C^2$ is used to represent $[\kappa_1,\kappa_2]$ sets of complex pairs.

The linear transformation is represented by a complex matrix:

$$\begin{bmatrix} \kappa_1 \\ \kappa_2 \end{bmatrix}\mapsto\begin{bmatrix} m_1 \\ m_2 \end{bmatrix}=\begin{bmatrix} a & b \\ c & d \end{bmatrix}\begin{bmatrix} \kappa_1 \\ \kappa_2 \end{bmatrix}=\begin{bmatrix} a\kappa_1+b\kappa_2 \\ c\kappa_1+d\kappa_2 \end{bmatrix}$$

()

If A $[m_1,m_2]$ and B $[\kappa_1,\kappa_2]$ are the $C^2$ homogeneous coordinates of point $z=(\kappa_1/\kappa_2)$ and its pixel $w=(m_1/m_2)$ in the complex set, respectively, then the linear transformation in $C^2$ is a nonlinear transformation in the complex set:

$$z = \frac{\kappa_1}{\kappa_2} \mapsto w = \frac{m_1}{m_2} = \frac{a\kappa_1 + b\kappa_2}{c\kappa_1 + d\kappa_2} = \frac{a\left(\kappa_1/\kappa_2\right) + b}{c\left(\kappa_1/\kappa_2\right) + d} = \frac{az + b}{cz + d}$$

(9)

The most general rotation of a Riemannian sphere can be expressed as a Mobius transformation

$$R(z) = \frac{az + b}{-\bar{b}z + \bar{a}},$$

(10)

which can be considered as mapping every point on the complex plane to a unique point on the Riemannian sphere.

To visualize the Mobius transformation $M(z) = \frac{az + b}{cz + d}$, we assume it has two fixed points $\xi_{_}$ and consider it as a mapping of itself, $z \mapsto \omega = M(z)$, as shown in **Fig. 2a**.

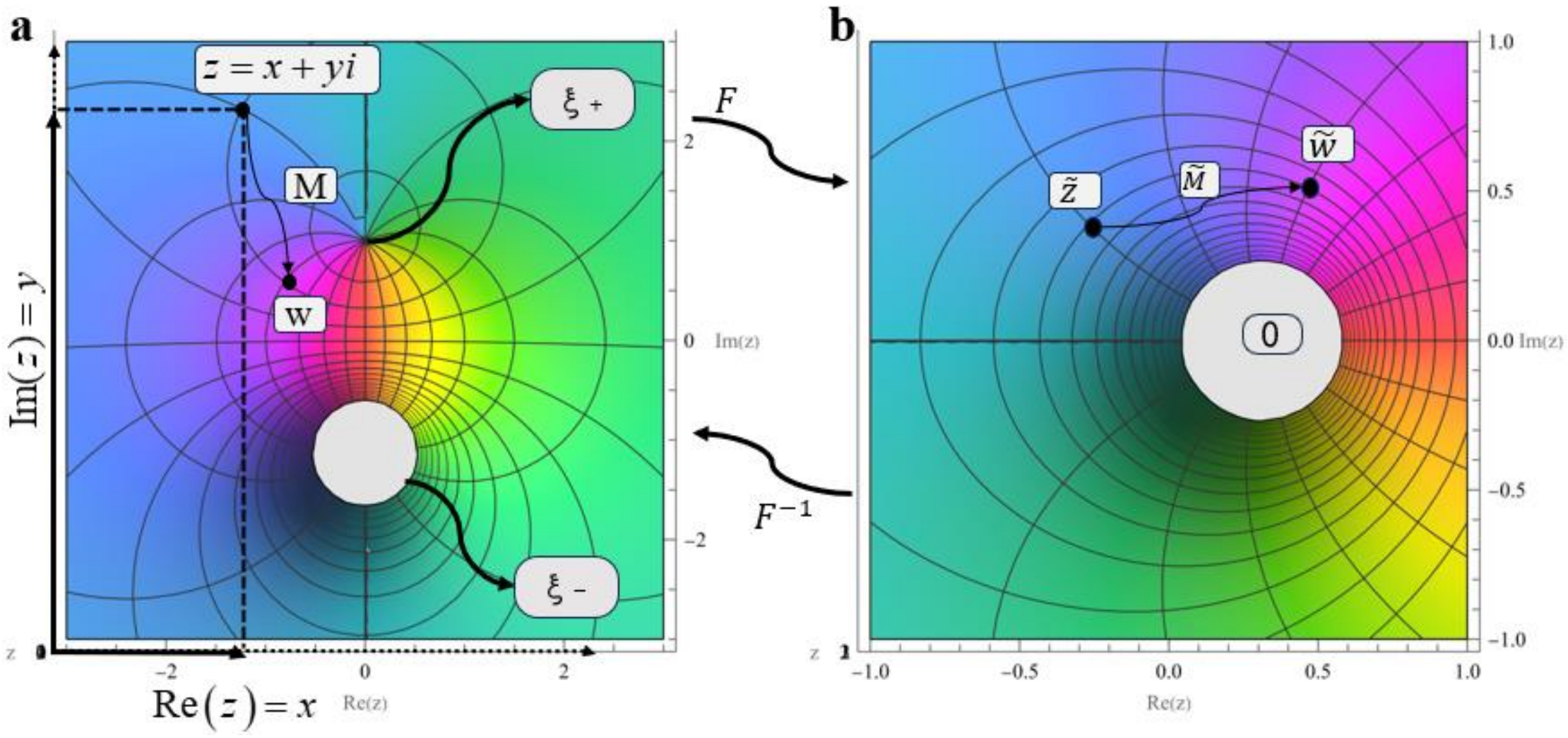

**Fig. 2** Mobius transformation diagram

In the simplest Mobius transformation (see **Fig. 2a**), $F(z) = \frac{z - \xi_+}{z - \xi_-}$ maps one fixed point (set to $\xi_+$) to zero and the other fixed point ($\xi_-$) to $\infty$. **Fig. 2b** illustrates the image on the left under this Mobius transformation.

## 3. Results

### 3.1 Geometric structures of $Ba_2LuNbO_6$ and $Ba_2ScNbO_6$

The molecular formula of perovskite-type oxides is $ABO_3$, where A is a larger cation and B is a smaller cation[11]. In the formed octahedral framework depicted in **Fig. 3**, the A cation occupies the central position while the B cation forms a 6-fold coordination with oxygen ions. Additionally, the A cation coordinates with oxygen ions in a 12-fold coordination. This structure is considered as a B-site cation occupying the center of the octahedron while the A-site cation is located at the center of the cube. Therefore, the perovskite structure is a superstructure formed by incorporating A-site cations into the $BO_6$ octahedron to form an $ABO_3$-type skeleton, which serves as the host structure in numerous compounds.

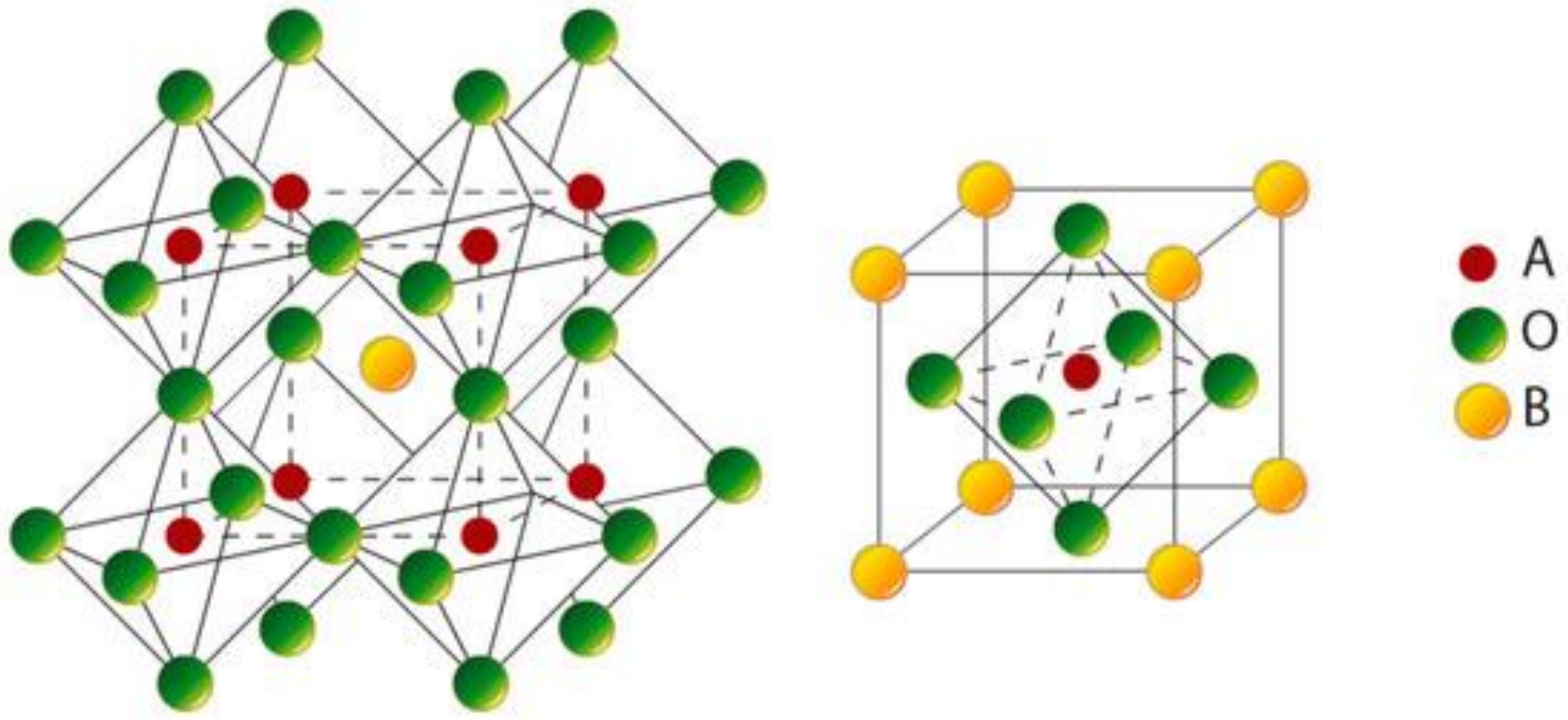

**Fig. 3** Geometric structures of $ABO_3$ perovskite

The molecular formula of the double perovskite compound is $A_2BB'O_6$. Its crystal structure can be expressed as a three-dimensional network where $BO_6$ and $B'O_6$ alternate to form an octahedron, whose interstitial positions are occupied by A atoms. In the standard perovskite $ABO_3$ structure, the B site—originally occupied by only one atom—is alternately occupied by two different atoms. As shown in the structural diagrams in Fig. 3, in the structures of $Ba_2LuNbO_6$, $LuO_6$, and $NbO_6$ alternate to form octahedra, whose interstices are filled by Ba. Similarly, in the other four compounds of $Ba_2ScNbO_6$, the gaps of the octahedra formed alternately by $BO_6$ (B: Lu, Sc) and $B'O_6$ (B': Nb) are filled with A-site Ba. The simulated X-ray powder diffraction (XRD) patterns of the two types of double perovskites (i.e., $Ba_2LuNbO_6$ and $Ba_2ScNbO_6$; see the **Supporting Information**) were roughly consistent with those in the database. Moreover, molecular dynamics simulations revealed that constructed crystal structure was stable.

HSE06 was also used for band structure and data analyses. The initially calculated geometric structure is shown in **Fig. 4,** and the lattice structure parameter sets for $Ba_2LuNbO_6$ and $Ba_2ScNbO_6$ are listed in **Table 2** and **Table 3**, respectively.

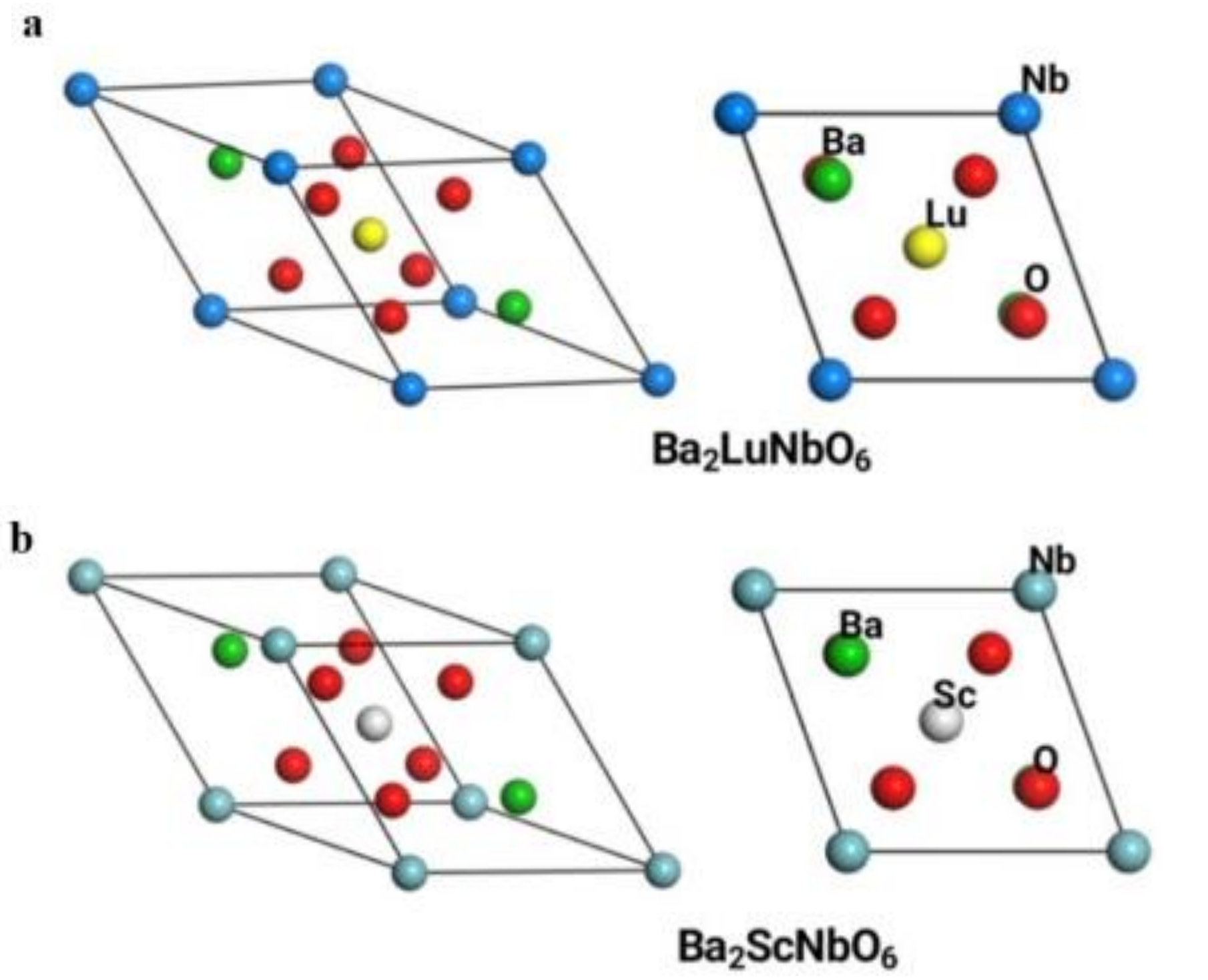

**Fig. 4** Geometrical structures of (a) $Ba_2LuNbO_6$ and (b) $Ba_2ScNbO_6$

**Table 2** Lattice and structural parameters of $Ba_2LuNbO_6$

| **Lattice parameters** | | | | | |
|---|---|---|---|---|---|
| **a** | b | c | α | β | γ |
| **5.85** | 5.85 | 5.85 | 60.00 | 60.00 | 60.00 |
| **Structure parameters** | | | | | |
| **Number** | Element | Atom | X | Y | Z |
| **1** | Ba | Ba（0） | 0.75 | 0.75 | 0.75 |
| **2** | Ba | Ba（1） | 0.25 | 0.25 | 0.25 |
| **3** | Lu | Lu（2） | 0.50 | 0.50 | 0.50 |
| **4** | Nb | Nb（3） | 0.00 | -0.00 | 0.00 |
| **5** | O | O（4） | 0.76 | 0.24 | 0.76 |
| **6** | O | O（5） | 0.24 | 0.24 | 0.76 |
| **7** | O | O（6） | 0.76 | 0.76 | 0.24 |
| **8** | O | O（7） | 0.76 | 0.24 | 0.24 |
| **9** | O | O（8） | 0.24 | 0.76 | 0.24 |
| **10** | O | O（9） | 0.24 | 0.76 | 0.76 |

**Table 3** Lattice and structural parameters of $Ba_2ScNbO_6$

| **Lattice parameters** | | | | | |
|---|---|---|---|---|---|
| **a** | b | c | α | β | γ |
| **5.87** | 5.87 | 5.87 | 59.98 | 59.98 | 59.98 |
| **Structure parameters** | | | | | |
| **Number** | Element | Atom | X | Y | Z |
| **1** | Ba | Ba（0） | 0.25 | 0.25 | 0.25 |
| **2** | Ba | Ba（1） | 0.75 | 0.75 | 0.75 |
| **3** | Sc | Sc（2） | 0.50 | 0.50 | 0.50 |
| **4** | Nb | Nb（3） | -0.00 | 0.00 | 0.00 |
| **5** | O | O（4） | 0.76 | 0.24 | 0.24 |
| **6** | O | O（5） | 0.24 | 0.76 | 0.76 |
| **7** | O | O（6） | 0.24 | 0.76 | 0.24 |
| **8** | O | O（7） | 0.76 | 0.24 | 0.76 |
| **9** | O | O（8） | 0.24 | 0.24 | 0.76 |
| **10** | O | O（9） | 0.76 | 0.76 | 0.24 |

## 3.2 Band structure, local density of states (LDOS), deformation charge density, and electronic properties of $Ba_2LuNbO_6$ and $Ba_2ScNbO_6$

DFT calculations of the band structure, DOS, and deformation charge density of materials are comprehensive. By analyzing the results, we can further investigate and control the electronic and chemical bonding properties of the materials, thereby better leveraging them. After constructing the crystal structures of $Ba_2LuNbO_6$ and $Ba_2ScNbO_6$, we obtained their corresponding band structures, LODS, and deformation charge densities, as shown in **Fig. 5**, **Fig. 6,** and **Fig. 7**, respectively.

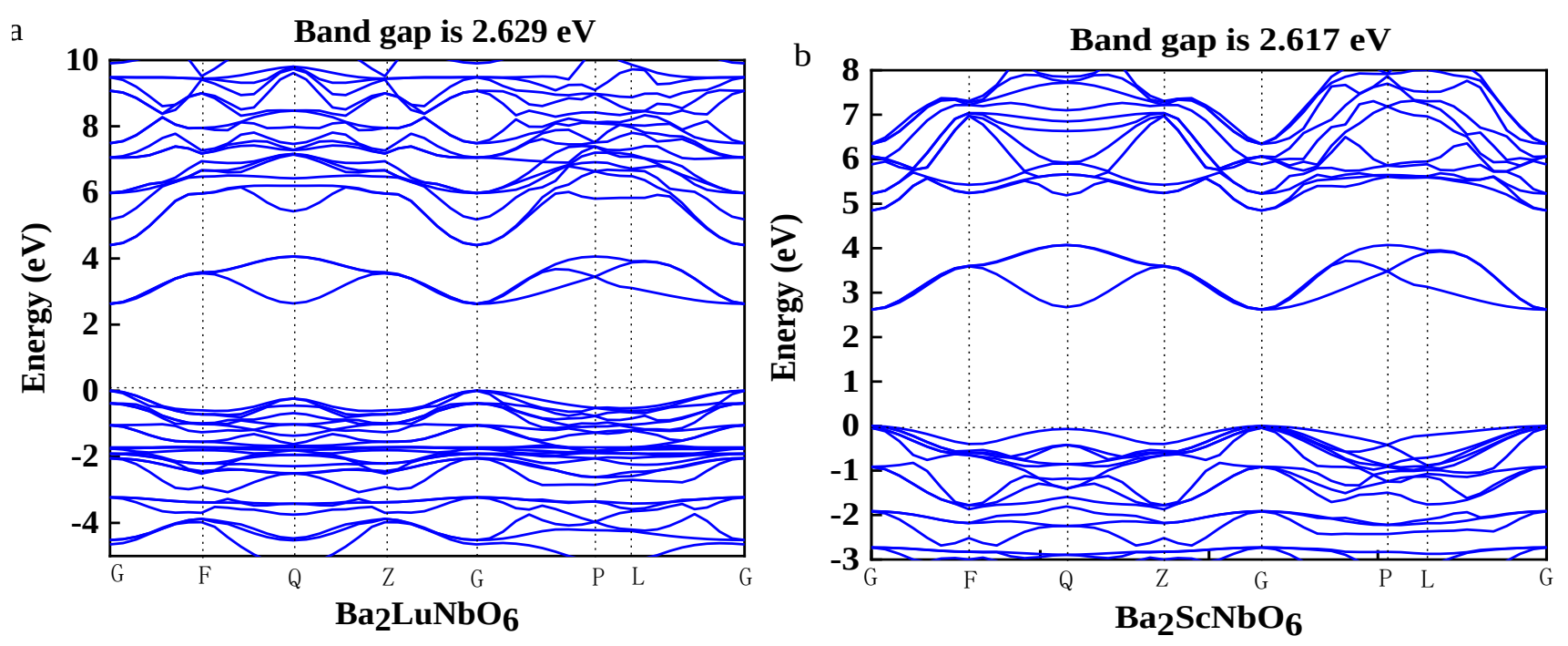

**Fig. 5** Band structure of (a) $Ba_2LuNbO_6$ and (b) $Ba_2ScNbO_6$

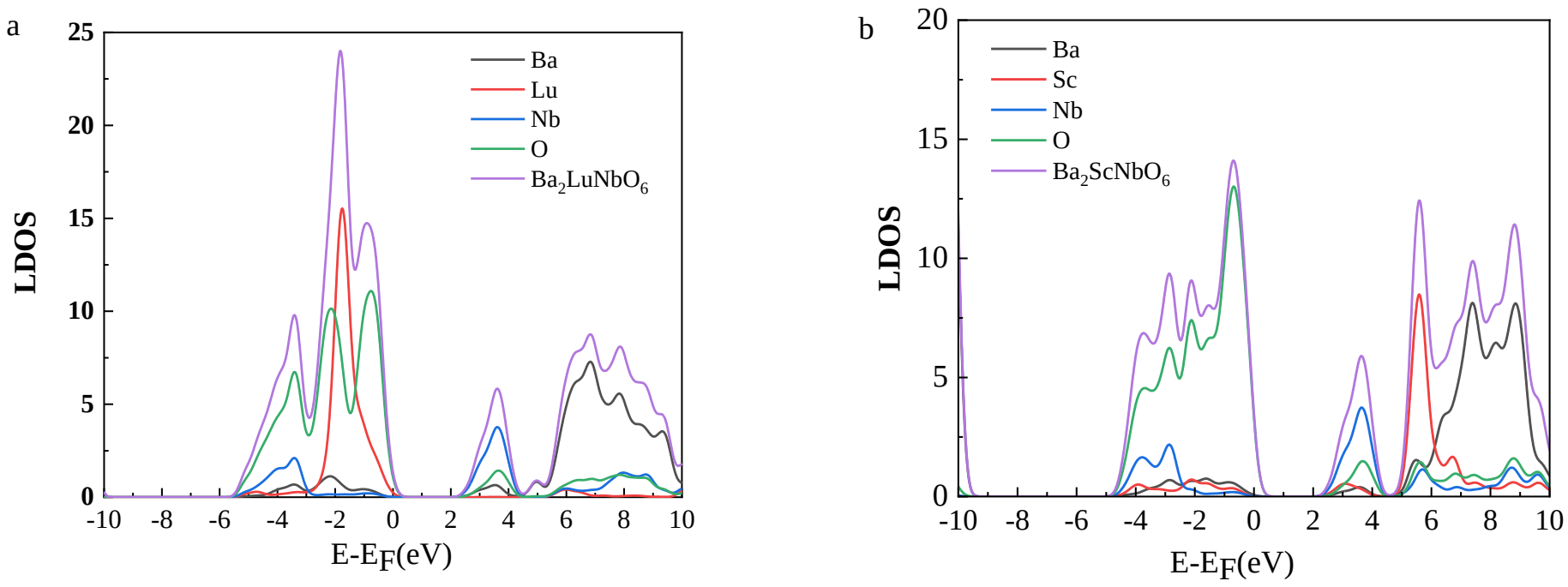

**Fig. 6** LDOS of (a) $Ba_2LuNbO_6$ and (b) $Ba_2ScNbO_6$

As shown in **Fig. 5**, $Ba_2LuNbO_6$ and $Ba_2ScNbO_6$ exhibit bandgaps of 2.629 eV and 2.617 eV, respectively, and are almost entirely occupied by electrons below the Fermi level. This observation, combined with the LDOS (**Fig. 6)**, indicate that both types of perovskites have semiconductor properties. Moreover, the lowest point of the conduction band and the highest point of the valence band of these two oxides are both at the same *k*-point, indicating that their positions in the *k*-space remain unchanged before and after electron transitions; therefore, they are both direct-bandgap semiconductors. However, there is a change in energy before and after electron transition, indicating a high probability of energy being released in the form of photons. Therefore, these two types of double perovskites can be used as alternative materials for the preparation of optical devices.

To analyze the electronic properties of the bonds, we used the deformation charge density (**Fig. 7**) of the O-Nb and O-Lu chemical bonds in $Ba_2LuNbO_6$ (**Fig. 7a,b**) and of the O-Nb and O-Sc chemical bonds in $Ba_2ScNbO_6$ (**Fig. 7c,d**). The electron distributions are indicated by the color scale, where the blue and red areas represent an increase and decrease, respectively, in the number of electrons. The O atom is mainly displayed in blue, indicating a positive deformation charge density and the accumulation of a large amount of charge in this region of the structure. The metal atoms of Lu, Nb, and Sc are mainly displayed in red, indicating a negative deformation charge density and a large amount of charge divergence.

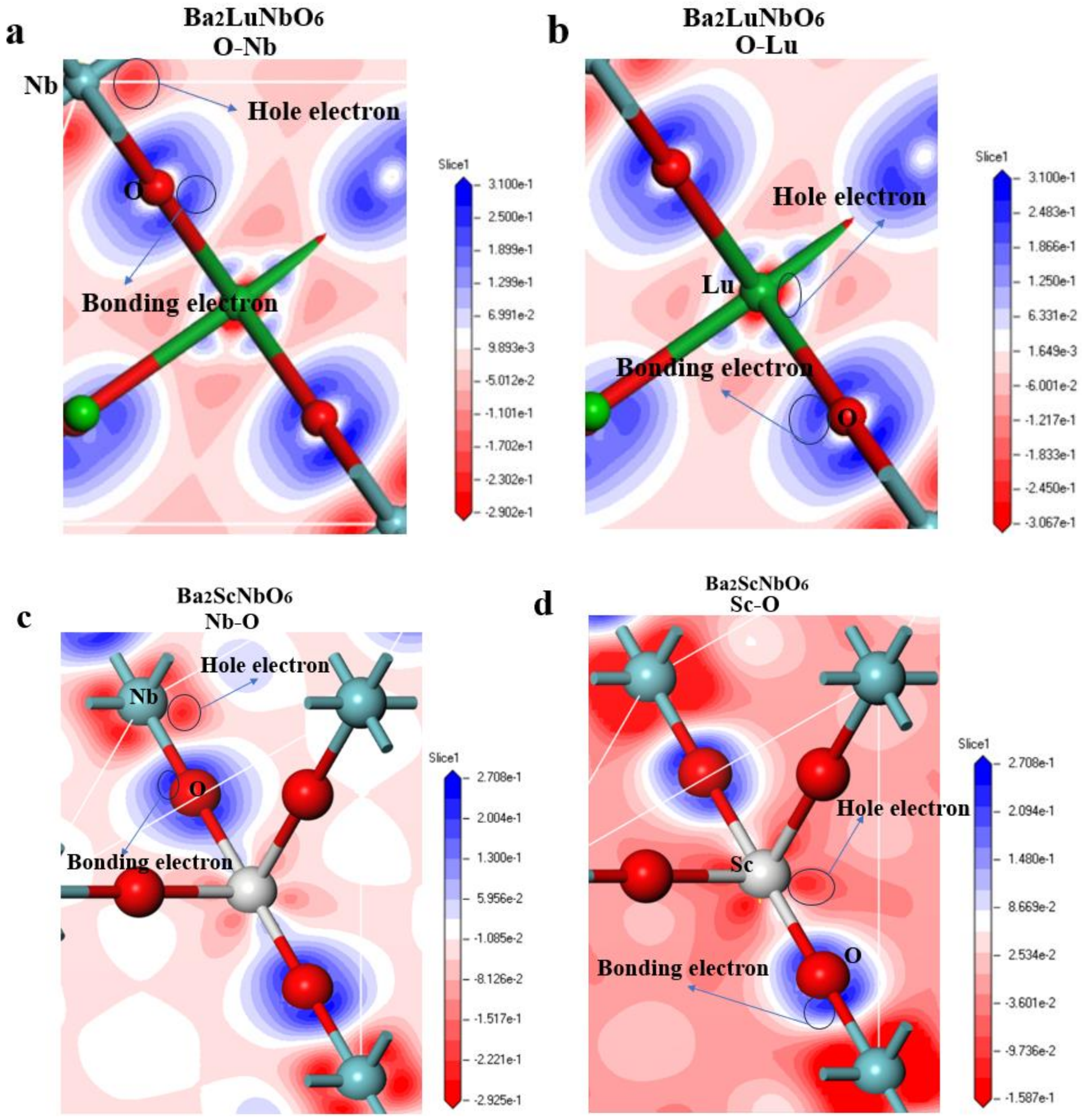

**Fig. 7** Deformation charge densities of $Ba_2LuNbO_6$ and $Ba_2ScNbO_6$

The established BBC model was used to convert the Hamiltonian values into bonding values (**Table 4**).

**Table 4** Deformation charge density $\delta\rho(\vec{r_{ij}})$ and deformation bond energy $\delta V_{bc}(\vec{r_{ij}})$, as obtained from the BBC model $\left(\varepsilon_0 = 8.85\times10^{-12}C^2N^{-1}m^{-2}, e = 1.60\times10^{-19}C, |\vec{r_{ij}}| \approx d_{ij}/2\right)$

| | $Ba_2LuNbO_6$ | $Ba_2LuNbO_6$ | $Ba_2ScNbO_6$ | $Ba_2ScNbO_6$ |
|---|---|---|---|---|
| $r_{ij}$ ( Å ) | 0.99(O-Nb) | 1.08（O-Lu） | 1.01(O-Nb) | 1.07(O-) |
| $r_i$ ( Å ) | 0.66（O） | 0.66（O） | 0.66（O） | 0.66（O） |
| $\vec{r_j}$ ( Å ) | 1.48(Nb) | 1.73(Lu) | 1.48(Nb) | 1.64(Sc) |
| $\delta\rho^{Hole-electron}\left(\overrightarrow{r_{ij}}\right)\left(e/\mathring{A}^3\right)$ | -0.2900 | -0.3067 | -0.2925 | -0.1587 |
| $\delta\rho^{Bonding-electron}\left(\overrightarrow{r_{ij}}\right)\left(e/\mathring{A}^3\right)$ | 0.3100 | 0.3100 | 0.2708 | 0.2708 |
| $\delta V_{bc}^{bonding}\left(\overrightarrow{r_{ij}}\right)(\mathrm{eV})$ | -0.6091 | -0.9420 | -0.5271 | -0.3672 |

The deformation charge density and electronic radius were used to calculate the deformation bond energies from **Eq. 4** as follows: -0.6091, -0.9420, -0.5271, and -0.3672 eV for O-Nb in $Ba_2LuNbO_6$, O-Lu in $Ba_2LuNbO_6$, O-Nb in$Ba_2ScNbO_6$ and O-Sc in $Ba_2ScNbO_6$, respectively.

### 3.3 Non-Hermitian bonding of $Ba_2LuNbO_6$ and $Ba_2ScNbO_6$

$$Z = \frac{az+b}{cz+d} = \frac{wz+b}{wz+d} \tag{11}$$

The Mobius transform is a fractional linear function that maps point *z* on a complex plane to a unique point, where *a*, *b*, *c*, and *d* are plurals and $ad-bc \neq 0$.

To use Eq. 11 for a point *W* = (*X*, *Y*, *Z*) in real space, we first normalize the coordinates $R_1(x_1, y_1, z_1)$ and $R_2(x_2, y_2, z_2)$ of two atoms in real space. The distance *l* between them is:

$$\sqrt{(x_1-x_2)^2+(y_1-y_2)^2+(z_1-z_2)^2} = l$$

Therefore:

$$\sqrt{\frac{(x_1-x_2)^2}{l^2}+\frac{(y_1-y_2)^2}{l^2}+\frac{(z_1-z_2)^2}{l^2}} = 1$$

$$\text{Order: } \begin{cases} X = \dfrac{x_1-x_2}{l} \\ Y = \dfrac{y_1-y_2}{l} \\ Z = \dfrac{z_1-z_2}{l} \end{cases}$$

Point *W* = (*X*, *Y*, *Z*) in real space satisfies $\sqrt{X^2+Y^2+Z^2} = 1$.

For a point $w$ in the complex number set $C$, $w = x + iy$, $\mathrm{Re}(w) = x$, and $\mathrm{Im}(w) = y$. As the projection point of $W$, $w$ satisfies $x = \frac{X}{1-Z}$ or $y = \frac{Y}{1-Z}$. The projections of the atomic coordinates of real space onto a two-dimensional complex plane are listed in **Table 5**.

**Table 5** Spatial coordinates of chemical bonds between $Ba_2ScNbO_6$ and $Ba_2LuNbO_6$ and their projection points on the complex plane

| | $Ba_2LuNbO_6$ | $Ba_2LuNbO_6$ | $Ba_2ScNbO_6$ | $Ba_2ScNbO_6$ |
|---|---|---|---|---|
| Element (coordinates) | O (1.14,4.26,4.33) | O (1.14, 4.26,4.33) | O (1.16, 4.26, 4.36) | O (1.16,4.26,4.36) |
| Element (coordinates) | Nb (0, 5.07, 2.93) | Lu (2.39,3.38,5.85) | Nb (0, 5.08, 2.94) | Sc (2.39, 3.39, 5.87) |
| $X$ | 0.5773 | 0.5773 | 0.5773 | 0.5773 |
| $Y$ | 0.4082 | 0.4082 | 0.4082 | 0.4081 |
| $Z$ | 0.7071 | 0.7071 | 0.7072 | 0.7072 |
| $W$ | 1.9712+1.3939$i$ | 1.9712+1.3939$i$ | 1.9716+1.3940$i$ | 1.9717+1.3940$i$ |

Based on **Eq. 5**, the projection points of O-Nb in $Ba_2LuNbO_6$, O-Lu in $Ba_2LuNbO_6$, O-Nb in $Ba_2ScNbO_6$, and O-Sc in $Ba_2ScNbO_6$ in a two-dimensional complex plane are $w$=1.992+1.3939$i$, $w$=1.9712+1.3939$i$, $w$=1.9716+1.3940$i$, and $w$=1.9717+1.3940$i$, respectively.

The elements of the matrix $\begin{bmatrix} a & b \\ c & d \end{bmatrix}$ in the Mobius transformation are complex numbers. To determine the projection of chemical bond energy, according to **Table 4**, we define the bond energy coefficients $V_{11}$, $V_{22}$, and $V_{12}$ as:

$$\begin{cases} V_{11} = \frac{1}{4\pi\varepsilon_0} \frac{e^2}{2|\vec{r}_j|} \int d^3r \int d^3r \delta\rho^{Hole-electron}(\vec{r}_j) \delta\rho^{Hole-electron}(\vec{r}_j) = \delta V_{bc}^{Antibonding}(\vec{r}_j) \\ V_{22} = \frac{1}{4\pi\varepsilon_0} \frac{e^2}{2|\vec{r}_i|} \int d^3r \int d^3r \delta\rho^{Bonding-electron}(\vec{r}_i) \delta\rho^{Bonding-electron}(\vec{r}_i) = \delta V_{bc}^{Antibonding}(\vec{r}_i) \\ V_{12} = \frac{1}{4\pi\varepsilon_0} \frac{e^2}{2|\vec{r}_i - \vec{r}_j|} \int d^3r \int d^3r \delta\rho^{Bonding-electron}(\vec{r}_{ij}) \delta\rho^{Hole-electron}(\vec{r}_{ij}) = \delta V_{bc}^{bonding}(\vec{r}_{ij}) \end{cases}$$

(12)

Substituting Eq. (12) and $w=x+iy$ into the matrix elements of the Mobius

transformation yields:

$$\begin{bmatrix} a & b \\ c & d \end{bmatrix} = \begin{bmatrix} w & -V_{11}i \\ w & -V_{22}i \end{bmatrix}$$

(13)

Substituting Eq. (13) into Eq. (11) and solving assuming *a=c=w*, we obtain:

$$V_{22} = \delta V_{bc}^{Antibonding}\left(\vec{r_i}\right) V_{11} = \delta V_{bc}^{Antibonding}\left(\vec{r_j}\right), z = V_{12} = V_{21} = \left|\delta V_{bc}^{bonding}\left(\vec{r_{ij}}\right)\right|$$

(14)

Substituting Eq. (14) into Eq. (11) yields the Mobius transformation with the chemical bond energy as a coefficient:

$$F(z) = \frac{V_{11}i - V_{12}w}{V_{22}i - V_{12}w}$$

(15)

The chemical bond energy as a coefficient is used to determine the eigenvalues of the Hamiltonian energy, which are transformed into energy projections in the complex space through the Mobius transformation. **Table 6** lists the Mobius transformation of the chemical bond energy coefficients of $Ba_2LuNbO_6$ and $Ba_2ScNbO_6$ computed by substituting **Eq. 12** into **Eq. 15**.

**Table 6** Mobius transformation of chemical bond energy *F*(*z*)

| | Bond | *F*(*z*) | *w* |
|---|---|---|---|
| $Ba_2LuNbO_6$ | O-Nb | $F(z) = \frac{4.2991i - 0.6091w}{0.0865i - 0.6091w}$ | $w = 1.9712 + 1.3939i$ |
| $Ba_2LuNbO_6$ | O-Lu | $F(z) = \frac{10.4786i - 0.9420w}{0.0865i - 0.9420w}$ | $w = 1.9712 + 1.3939i$ |
| $Ba_2ScNbO_6$ | O-Nb | $F(z) = \frac{4.3683i - 0.5271w}{0.0660i - 0.5271w}$ | $w = 1.9716 + 1.3940i$ |
| $Ba_2ScNbO_6$ | O-Sc | $F(z) = \frac{2.1489i - 0.3672w}{0.0660i - 0.3672w}$ | $w = 1.9717 + 1.3940i$ |

In order to obtain the various forms of chemical bonds in the complex space, we calculated the energy projection of two $Ba_2XYO_6$ chemical bonds using three functions, namely, *z=w, z=Zeta(w)*, and $z=e^w$, where the x-direction is the real part and the y-direction is the imaginary part (see **Fig. 8**). The O-Lu and O-Nb chemical bonds of $Ba_2LuNbO_6$, and the planar coordinates of the O-Sc and O-Nb chemical bonds of $Ba_2ScNbO_6$, were projected onto a Riemannian sphere using a complex energy Mobius transformation to obtain the non-Hermitian chemical bond energy projection of the chemical bonds in $Ba_2ScNbO_6$ and $Ba_2LuNbO_6$.

The calculations using the *z=w* function, illustrated in **Fig. 8a**, are as follows:

$$F(z)=\frac{4.2991i-0.6091w}{0.0865i-0.6091w}$$ , $w=1.9712+1.3939i$ of O-Nb in $Ba_2LuNbO_6$;

$$F(z)=\frac{10.4786i-0.9420w}{0.0865i-0.9420w}$$ , $w=1.9712+1.3939i$ of O-Lu in $Ba_2LuNbO_6$;

$$F(z)=\frac{4.3683i-0.5271w}{0.0660i-0.5271w}$$ , $w=1.9716+1.3940i$ of O-Nb in $Ba_2ScNbO_6$ ;

$$F(z)=\frac{2.1489i-0.3672w}{0.0660i-0.3672w}$$ , $w=1.9717+1.3940i$ of O-Sc in $Ba_2ScNbO_6$.

The calculations using the *z=Zeta(w)* function, illustrated in **Fig. 8b**, are as follows:

$$F(z)=\frac{4.2991i-0.6091Zeta[w]}{0.0865i-0.6091Zeta[w]}$$ , $w=1.9712+1.3939i$ of O-Nb in $Ba_2LuNbO_6$;

$$F(z)=\frac{10.4786i-0.9420Zeta[w]}{0.0865i-0.9420Zeta[w]}$$ , $w=1.9712+1.3939i$ of O-Lu in $Ba_2LuNbO_6$;

$$F(z)=\frac{4.3683i-0.5271Zeta[w]}{0.0660i-0.5271Zeta[w]}$$ , $w=1.9716+1.3940i$ of O-Nb in $Ba_2ScNbO_6$ ;

$$F(z)=\frac{2.1489i-0.3672Zeta[w]}{0.0660i-0.3672Zeta[w]}$$ , $w=1.9717+1.3940i$ of O-Sc in $Ba_2ScNbO_6$.

The calculations using the $z=e^w$ function, illustrated in **Fig. 8c**, are as follows:

$$F(z)=\frac{4.2991i-0.6091e^w}{0.0865i-0.6091e^w}$$ , $w=1.9712+1.3939i$ of O-Nb in $Ba_2LuNbO_6$;

$$F(z)=\frac{10.4786i-0.9420e^w}{0.0865i-0.9420e^w}$$ , $w=1.9712+1.3939i$ of O-Lu in $Ba_2LuNbO_6$;

$$F(z)=\frac{4.3683i-0.5271e^w}{0.0660i-0.5271e^w}$$ , $w=1.9716+1.3940i$ of O-Nb in $Ba_2ScNbO_6$ ;

$$F(z)=\frac{2.1489i-0.3672e^w}{0.0660i-0.3672e^w}$$ , $w=1.9717+1.3940i$ of O-Sc in $Ba_2ScNbO_6$.

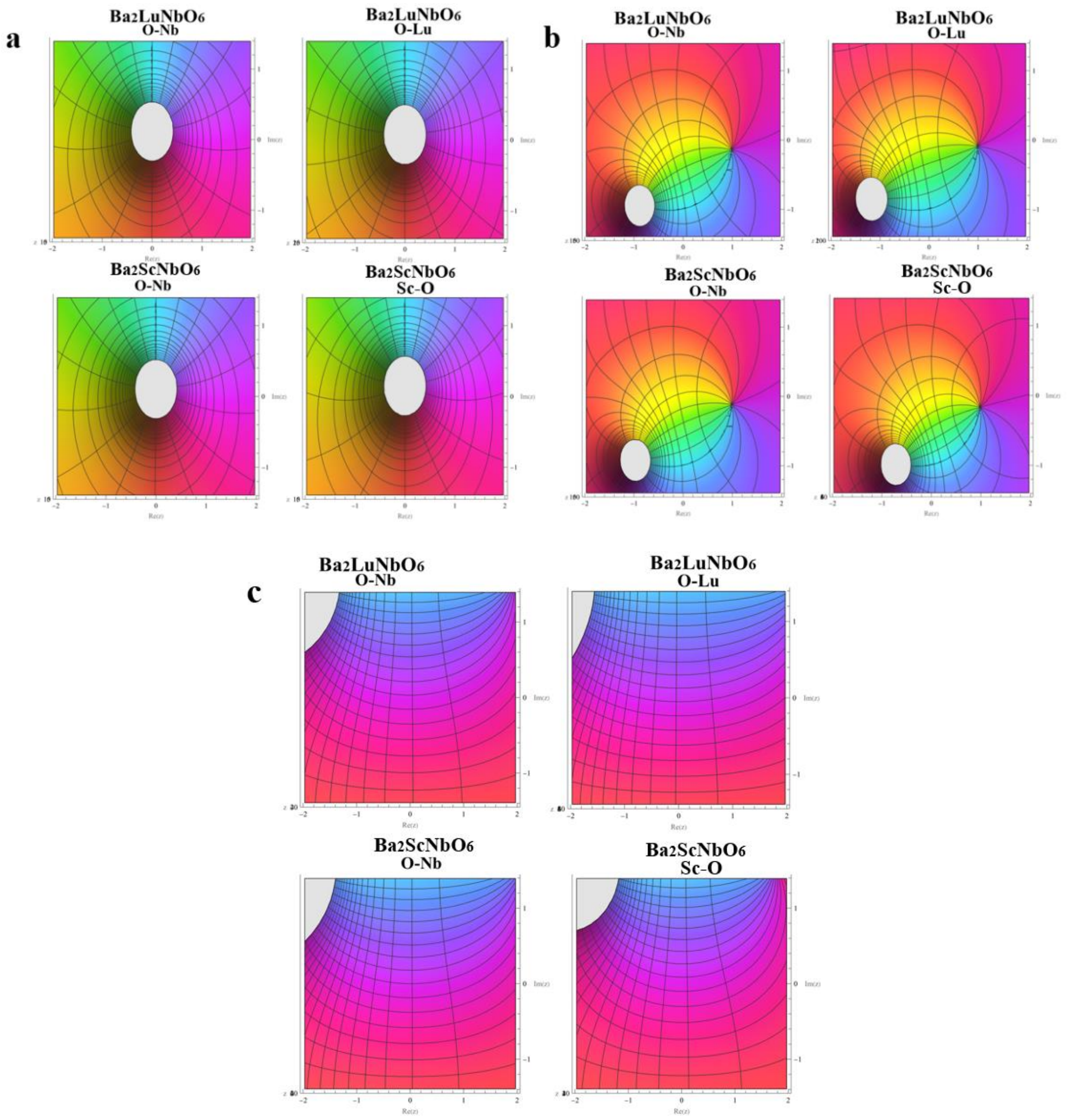

**Fig. 8** Non-Hermitian chemical bond energy projection of $Ba_2ScNbO_6$ and $Ba_2LuNbO_6$

In **Fig. 8**, singular zeros are observed in the projection plots of the $Ba_2XYO_6$ chemical bond energies calculated using the three different functions. The physical meaning of zero can be further explored. Considering the trajectory of particle motion in the system as the microscopic unit, we rely on the dynamic phase transition theory to explain anomalous diffusion behavior changes in phase diagrams that cannot be explained by traditional thermodynamics. The dynamic phase transition is strictly defined using the zero point of the dynamic ensemble. Early research emphasized that the observable properties are affected only when the zero point approaches the real axis (when phase transition occurs). However, the zero point essentially contains all the statistical information of the ensemble; thus, knowing the zero point is equivalent to knowing the partition function. This implies that we can have a more powerful Lee – Yang statistical theory than the current one, which could be formulated from a zero-point perspective. Peng et al. were the first to observe the Lee – Yang zero point of a spin system in their experiments[12]. They used the spin as a probe to connect with the system of the observed zero point. By analyzing the dynamic behavior of the probe, they converted complex spatial information onto the time axis, thereby determining the position of the zero point on the complex plane.

The non-Hermitian zeros of the O-Nb chemical bond in $Ba_2LuNbO_6$ and the non Hermitian zeros of the O-Lu chemical bond *z=Zeta(w)* and $z=e^w$ are inclusive relationships in *z=w* function and function expressions, and intersecting relationships in function expressions. The relationships between the non-Hermitian zeros of the O-Nb chemical bond of $Ba_2ScNbO_6$ and those of the Sc-O chemical bond are similar but vary in size. This interaction between the chemical bonds of the two types of $Ba_2XYO_6$ is reflected through the zero points, as shown in **Fig. 9**.

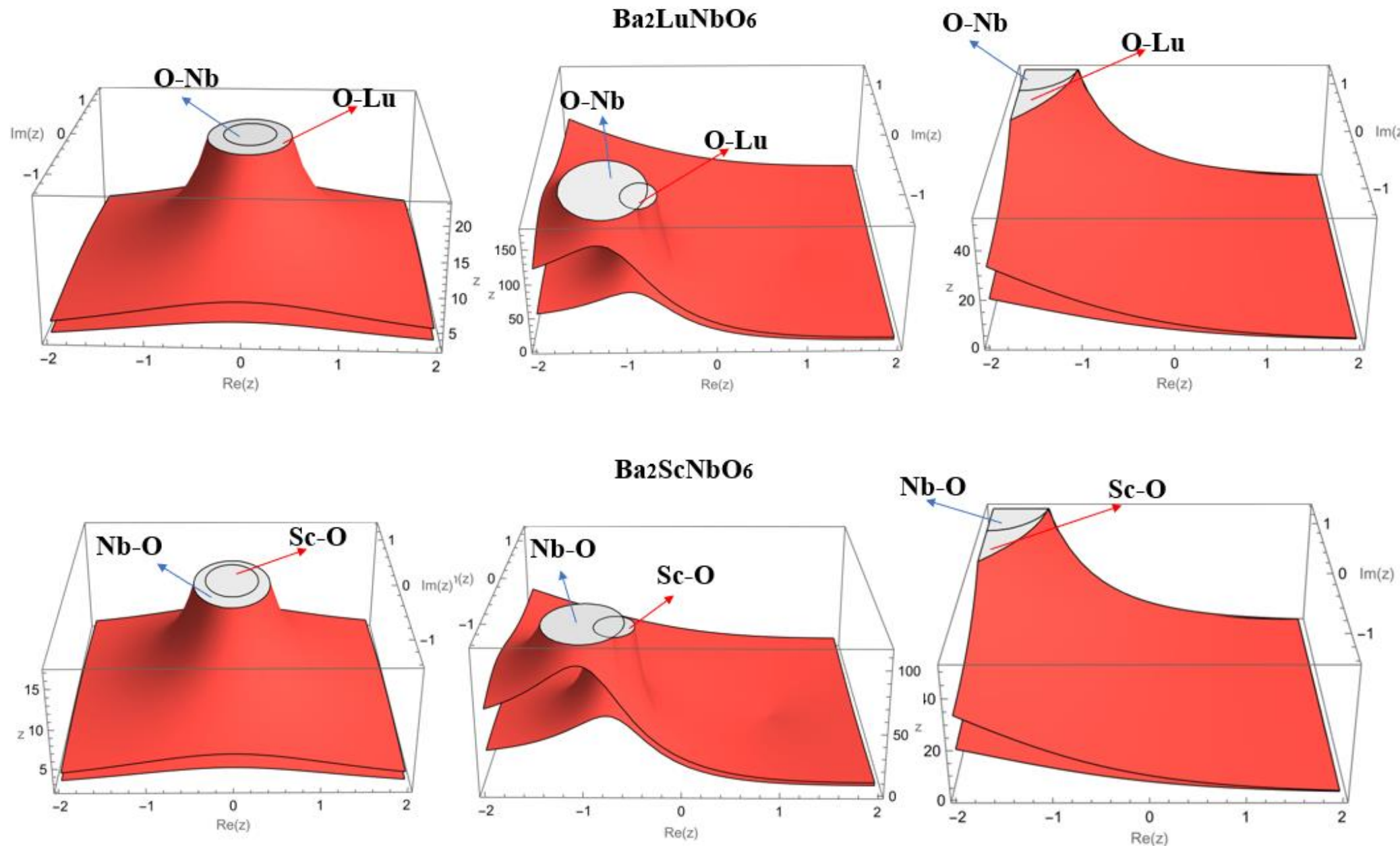

**Fig. 9** Non-Hermitian bonding of $Ba_2ScNbO_6$ and $Ba_2LuNbO_6$ reflected through the zero points

The abscissa and ordinate represent the real and imaginary parts of the complex field, respectively, and the *z* axis represents the height of the potential barrier. The chemical bonds projected in the Riemannian space through a Mobius transformation enable the detection of the zero-point position on the complex plane (**Fig. 9**). Based on the mathematical similarity, the zero point corresponds to the free energy as the charge corresponds to the potential. The zero-point distribution can be understood as the spatial distribution of charge, similar to electrostatic shielding and leakage phenomena, and explains the effect of the zero point on phase transitions and supercritical regions.

## 4. Conclusion

We calculated the electronic structures of $Ba_2ScNbO_6$ and $Ba_2LuNbO_6$ using DFT calculations and demonstrated their semiconductor properties by analyzing their band structures, DOS, and deformation charge densities. Further, using the BBC model, we calculated the differential bond energies of O-Nb in $Ba_2LuNbO_6$, O-Lu in $Ba_2LuNbO_6$, O-Nb in $Ba_2ScNbO_6$, and O-Sc in $Ba_2ScNbO_6$, yielding vales of -0.6091 eV, -0.9420 eV, -0.5271 eV, and -0.3672 eV, respectively. In addition, we calculated the energy projections of the non-Hermitian chemical bonds in $Ba_2ScNbO_6$ and $Ba_2LuNbO_6$. The non-Hermitian zero point corresponds to the free energy, as the charge corresponds to the potential. Our results provide a novel and effective method for discerning the electronic bonding properties of $Ba_2XYO_6$ oxides and lay the foundation for the application of non-Hermitian materials in various chemical studies.

# Support Information

## Non-Hermitian bonding and electronic reconfiguration of $Ba_2ScNbO_6$ and $Ba_2LuNbO_6$

Yaorui Tan, Maolin Bo*

Key Laboratory of Extraordinary Bond Engineering and Advanced Materials Technology (EBEAM) of Chongqing, Yangtze Normal University, Chongqing 408100, China

*Author to whom correspondence should be addressed: bmlwd@yznu.edu.cn (Maolin Bo)

## I. Hamiltonian

For bond-charge (BC) model, we consider the positive charge background (b) and the electron (e) as a system, and write their Hamiltonian sums and their interactions, respectively. In addition to electron kinetic energy, only electrostatic Coulomb interactions are considered:

$$H = H_b + H_e + H_{eb}$$

(1)

$$H_e = \sum_{i=1}^{N} \frac{P_i^2}{2m} + \frac{1}{2} e_1^2 \sum_{i=1}^{N} \sum_{\substack{j=1 \\ i \neq j}}^{N} \frac{1}{|r_i - r_j|} e^{-\mu |r_i - r_j|}$$

(2)

$$H_b = \frac{1}{2} e_1^2 \int \mathrm{d}^3 x \int \mathrm{d}^3 x' \frac{n(x) n(x')}{|x - x'|} e^{-\mu |x - x'|}$$

(3)

$$H_{eb} = -e_1^2 \sum_{i=1}^{N} \int \mathrm{d}^3 x \frac{n(x)}{|x - r_i|} e^{-\mu |x - r_i|} = -e_1^2 \sum_{i=1}^{N} \frac{N}{V} 4\pi \int \mathrm{d}z \frac{e^{-\mu z}}{z} = -4\pi e_1^2 \frac{N^2}{V \mu^2}$$

(4)

The shielding factor $\mu$ is added to the equation. $r_i$ represents the *ith* electronic position. $x$ represents the background position. $e$ is the basic charge, $e_1 = e / \sqrt{4\pi\varepsilon_0}$ . Since the coulomb force is a long-range force, while $N \to \infty$, $V \to \infty$ , $N/V$ remains unchanged, if the coulomb potential energy is calculated separately, $H_e$ 、 $H_b$ and $H_{eb}$ will diverge. In physically, when increasing the volume by keeping the electron density ($N/V$) constant, the average electron energy $H/N$ should be a constant. Therefore, the divergent terms in the $H_e$ , $H_b$ and $H_{eb}$ should be able to cancel each other out, in order to see this, We have added an attenuation factor $\mu$ to (2), (3), and (4), after the integration $L$ of the cube with a side length of $L \to \infty$ , And then order $\mu \to 0$, this is a commonly used technique in solid state physics calculations. $e_1 n(x)$ is the charge density at background $x$, and $n(x) = N/V$ is a constant.

$$H_b = \frac{1}{2} e_1^2 \left(\frac{N}{V}\right)^2 \int d^3x 4\pi \int dz \frac{e^{-\mu z}}{z} = 4\pi e_1^2 \frac{N^2}{2V\mu^2}$$

(5)

$$H_{eb} = -e_1^2 \sum_{i=1}^{N} \frac{N}{V} 4\pi \int dz \frac{e^{-\mu z}}{z} = -4\pi e_1^2 \frac{N^2}{V\mu^2}$$

(6)

$$\sum_{i=1}^{N} \frac{P^2}{2m} = \sum_{l'\sigma'} \sum_{l\sigma} a_{k_{l'}\sigma'}^{\dagger} \left\langle k_{l'}\sigma' \left| \frac{P^2}{2m} \right| k_l \sigma \right\rangle = \sum_{l\sigma} \frac{\hbar^2 k_l^2}{2m} a_{k_l\sigma}^{\dagger} a_{k_l\sigma}$$

(7)

In the above formula, use $k_l$ representative $k_{xl}$ 、 $k_{yl}$ and $k_{zl}$, $l = 0, \pm 1, \pm 2, \pm 3, \cdots$ for typographical convenience, we sometimes use $k$ and $k'$ in lieu of $k_l$ and $k_{l'}$, just remember the momentum $k$ take discrete values instead of continuous values. The second item in（2） is

$$\frac{1}{2} e_1^2 \sum_{l'\sigma'} \sum_{m'\lambda'} \sum_{l\sigma} \sum_{m\lambda} a_{k_{l'}\sigma'}^{?} a_{k_{m'}\lambda'}^{\dagger} (k_{l'}\sigma', k_{m'}\lambda' \left| \frac{e^{-\mu|r_1 - r_2|}}{|r_1 - r_2|} \right| k_l\sigma, k_m\lambda) a_{k_m\lambda} a_{k_l\sigma}$$

$$= \frac{1}{2} e_1^2 \sum_{l'} \sum_{m'} \sum_{l\sigma} \sum_{m\lambda} \delta_{\sigma\sigma'} \delta_{\lambda\lambda'} a_{k_{l'}}^{?} a_{k_{m'}}^{\dagger} (k_{l'} k_{m'} \left| \frac{e^{-\mu|r_1 - r_2|}}{|r_1 - r_2|} \right| k_l k_m) a_{k_m} a_{k_l}$$

(8)

The matrix elements are first calculated in the position representation. Insert two completeness relations in the matrix element in the above equation:

$$\int d^3 r^\alpha \int d^3 r^\beta \left| r^\alpha r^\beta \right\rangle \left\langle r^\alpha r^\beta \right| = 1$$

$$\int d^3 r^\gamma \int d^3 r^\delta \left| r^\gamma r^\delta \right\rangle \left\langle r^\gamma r^\delta \right| = 1$$

And

$$(k_{l'}k_{m'}\left|\frac{e^{-\mu|r_1-r_2|}}{|r_1-r_2|}\right|k_l k_m)=\frac{1}{V^3}\int d^3r^\alpha\int d^3r^\beta e^{-i(k_{l'}-k_l)r^\alpha}e^{-i(k_{m'}-k_m)r^\beta}\frac{e^{-\mu|r^\alpha-r^\beta|}}{|r^\alpha-r^\beta|}$$

$$=\delta\left(k_{l'}+k_{m'},k_l+k_m\right)\frac{1}{V}\int\frac{e^{-\mu r}}{r}e^{-i(k_{l'}-k_l)r}d^3r$$

$$=\delta\left(k_{l'}+k_{m'},k_l+k_m\right)\frac{2\pi}{V}\int re^{-\mu r}\left(\frac{2}{kr}\sin kr\right)dr$$

$$=\delta\left(k_{l'}+k_{m'},k_l+k_m\right)\frac{1}{V}\frac{4\pi}{\left(k_{l'}-k_l\right)^2+\mu^2}$$

（9）

In the above calculation, the second step uses the independent variable substitution： $r^\alpha+r^\beta=r'$ 、 $r^\alpha-r^\beta=r$ [Jacobian determinant $J=\left(1/2\right)^3$ ]，substitute this equation back（8）, $k_l=k$ , $k_m=k'$ , $k_{l'}=k+q$ , $k_{m'}=k'-q$ , the second item in $H_e$ is

$$\frac{e_1^2}{2V}\sum_k\sum_{k'}\sum_q\sum_{\sigma\lambda}\frac{4\pi}{q^2+\mu^2}a^\dagger_{k+q,\sigma}a^\dagger_{k'-q,\lambda}a_{k'\lambda}a_{k\sigma}$$

（10）

Separate out $q=0$ is

$$\frac{e_1^2}{2V}\sum_k\sum_{k'}\sum_q\sum_{\sigma\lambda}\frac{4\pi}{\mu^2}a^\dagger_{k\sigma}a^\dagger_{k'\lambda}a_{k'\lambda}a_{k\sigma}=\frac{e_1^2}{2V}\frac{4\pi}{\mu^2}\left(N^2-N\right)$$

*N* is the total number of particles，because

$$a^\dagger_{k\sigma}a^\dagger_{k'\lambda}a_{k'\lambda}a_{k\sigma}=N_{k\sigma}N_{k'\lambda}-N_{k\sigma}\delta_{kk'}\delta_{\sigma\lambda}$$

Now substitute (7) and (10) back to equation (2), $H_e$ is

$$H_e=\sum_{k\sigma}\frac{\hbar^2k^2}{2m}a^\dagger_{k\sigma}a_{k\sigma}+\frac{e_1^2}{2V}\sum_q{}^*\sum_{k\sigma}\sum_{k'\lambda}\frac{4\pi}{q^2+\mu^2}a^\dagger_{k+q,\sigma}a^\dagger_{k'-q,\lambda}a_{k'\lambda}a_{k\sigma}+\frac{e_1^2}{2V}\frac{4\pi}{\mu^2}\left(N^2-N\right)$$

（11）

In the formula, the *q* on the sum sign of "*" indicates that the part where q=0 is ignored during the sum. When $V\to\infty,N\to\infty$ , while keeping *N*/*V* constant, the last term to the right of the equal sign causes the average energy $H_e/N$ of each particle to become:

$$\frac{1}{2}4\pi e_1^2\left(\frac{N}{V}\right)\frac{1}{\mu^2}-\frac{1}{2}4\pi e_1^2\left(\frac{N}{V}\right)\frac{1}{N}\frac{1}{\mu^2}$$

The former term is constant, and the latter term tends to zero. If $\mu \to 0$, the former term becomes a divergent term. However, this term just cancels out the divergent $H_b$ and $H_{eb}$ term. Thus, the Hamiltonian of the system becomes

$$H=\sum_{k\sigma}\frac{\hbar^2k^2}{2m}a_{k\sigma}^{\dagger}a_{k\sigma}+\frac{e_1^2}{2V}\sum_{q}\sum_{k'}\sum_{\sigma\lambda}\frac{4\pi}{q^2}a_{k+q,\sigma}^{\dagger}a_{k'-q,\lambda}^{\dagger}a_{k'\lambda}a_{k\sigma}=H_0+H_1$$

(12)

## II. Structural stability

We used molecular dynamics (MD) to simulate the structural stability of bilayer $Ba_2ScNbO_6$ and $Ba_2LuBiO_6$ materials. The $Ba_2ScNbO_6$ and $Ba_2LuBiO_6$ structure were obtained by performing NVE ensemble with time increments at 1 fs for 100 ps (the total iteration steps are 100,000) until the potential energy accomplished a stable value. The box size, volume, number of atoms of $Ba_2ScNbO_6$ and $Ba_2LuBiO_6$ structure, as is shown in the Table S1. The molecular dynamics simulation results show that the structures of $Ba_2ScNbO_6$ and $Ba_2LuBiO_6$ are stable. The XRD diffraction patterns of $Ba_2ScNbO_6$ and Ba2LuBiO6 consistent with those in the database, so the crystal structure we constructed is reasonable.

**Table S1** Box size, volume and number of atoms of $Cd_{43}Te_{28}$ microporous materials.

| Structure | Number of atoms | Volume | Box size | | | Angle | | |
|---|---|---|---|---|---|---|---|---|
| | | | a(Å) | b(Å) | c(Å) | α | β | γ |
| Ba2ScNbO6 | 2160 | 30831.795 Å $^3$ | 35.202 | 35.202 | 35.202 | 59.982° | 59.982° | 59.982° |
| Ba2LuBiO6 | 2160 | 30629.191 Å$^3$ | 35.120 | 35.120 | 35.120 | 60.000° | 60.000° | 60.000° |

(1) Structural stability of $Ba_2ScNbO_6$

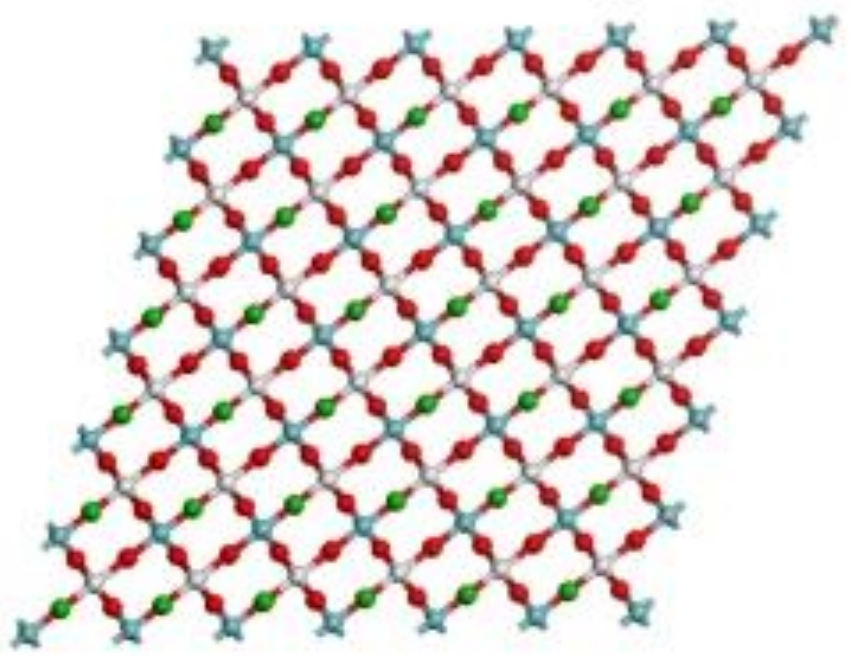

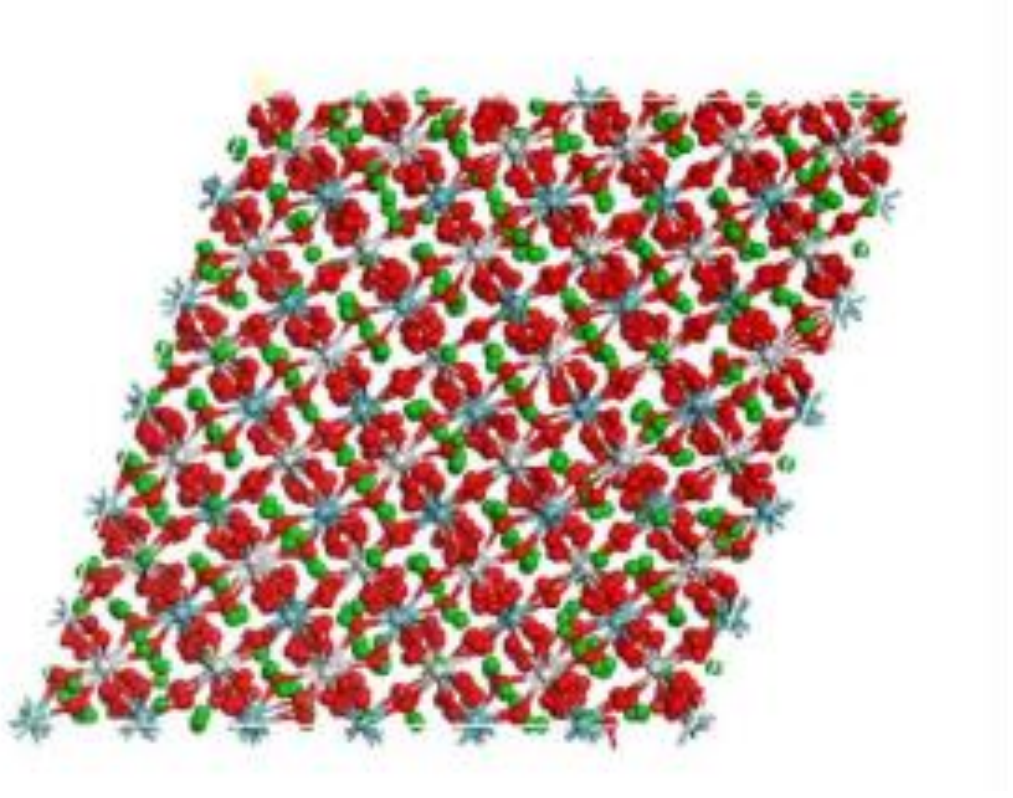

0K 298K

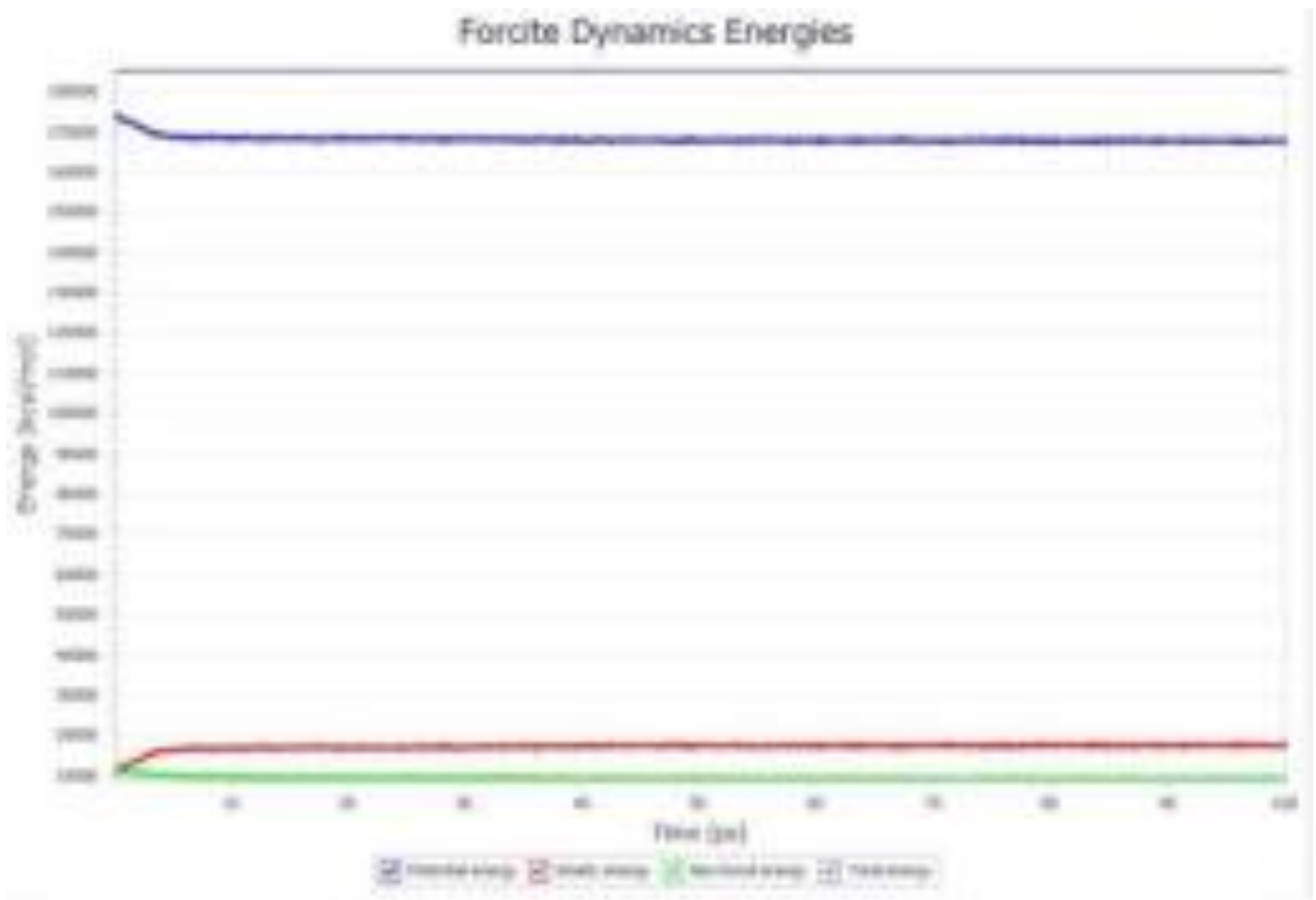

(2) Structural stability of $Ba_2LuBiO_6$

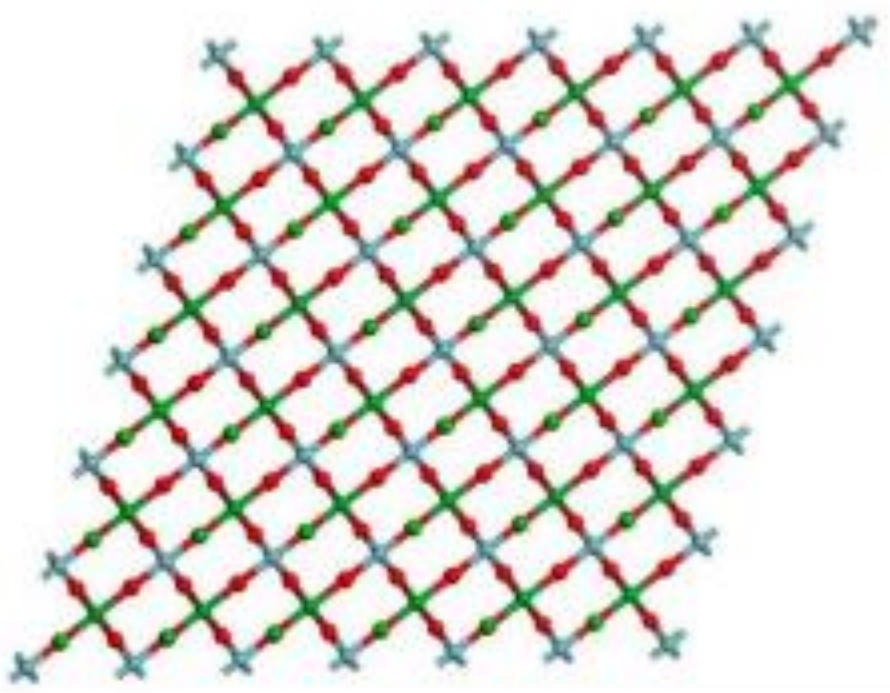

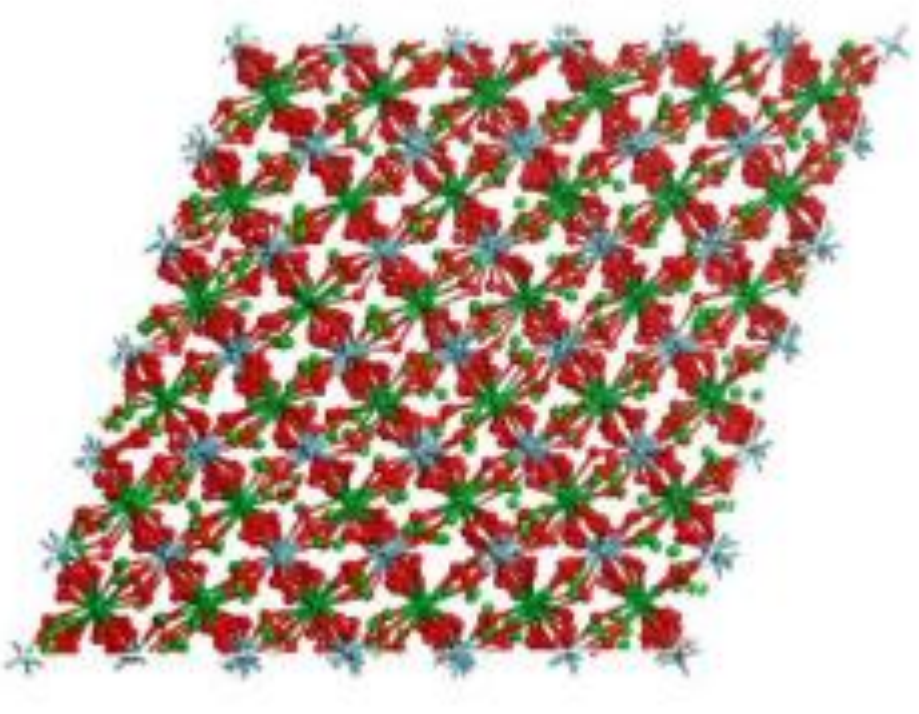

0K 298K

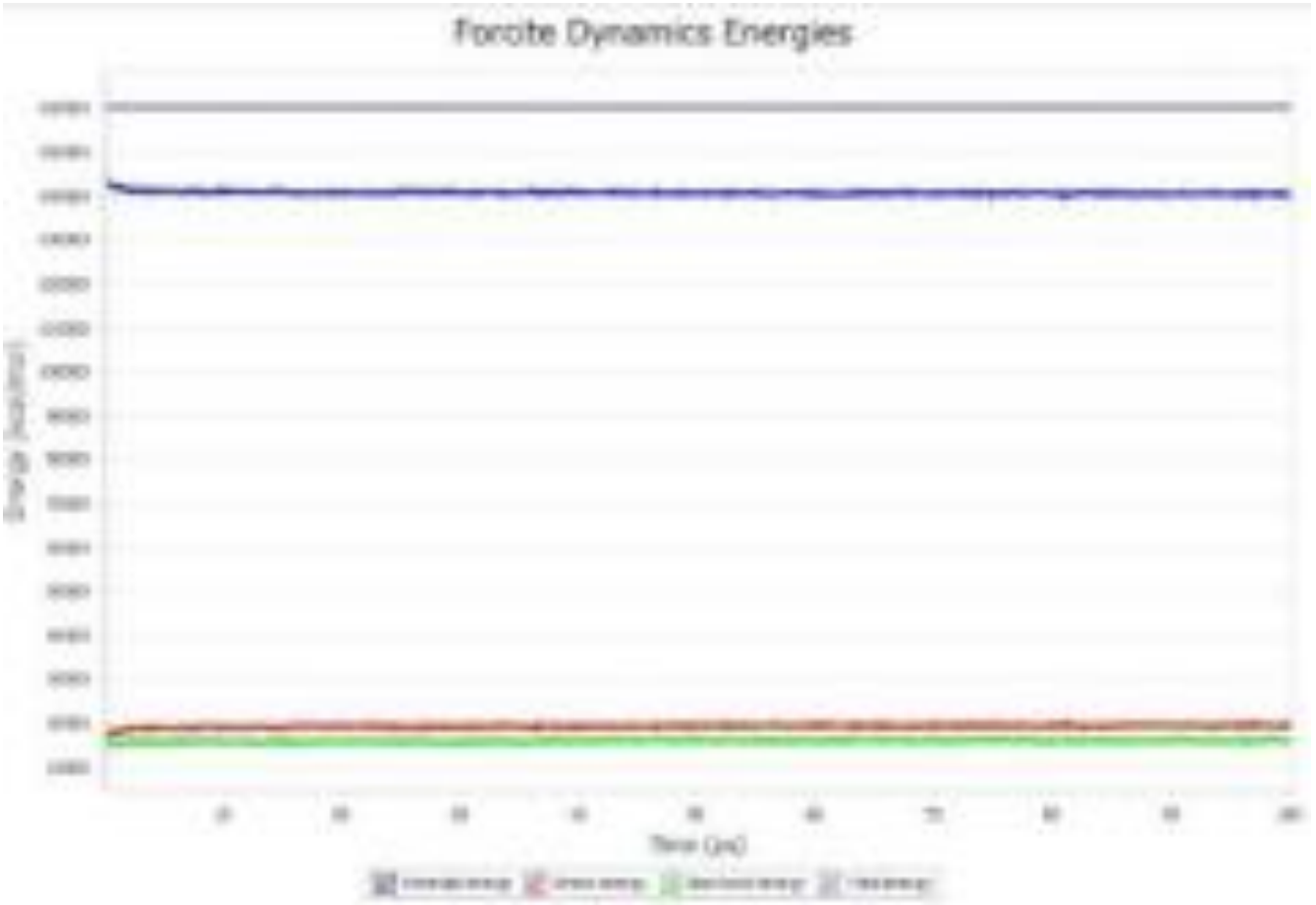

(3) XRD diffraction patterns of $Ba_2luNbO_6$ and $Ba_2ScNbO_6$

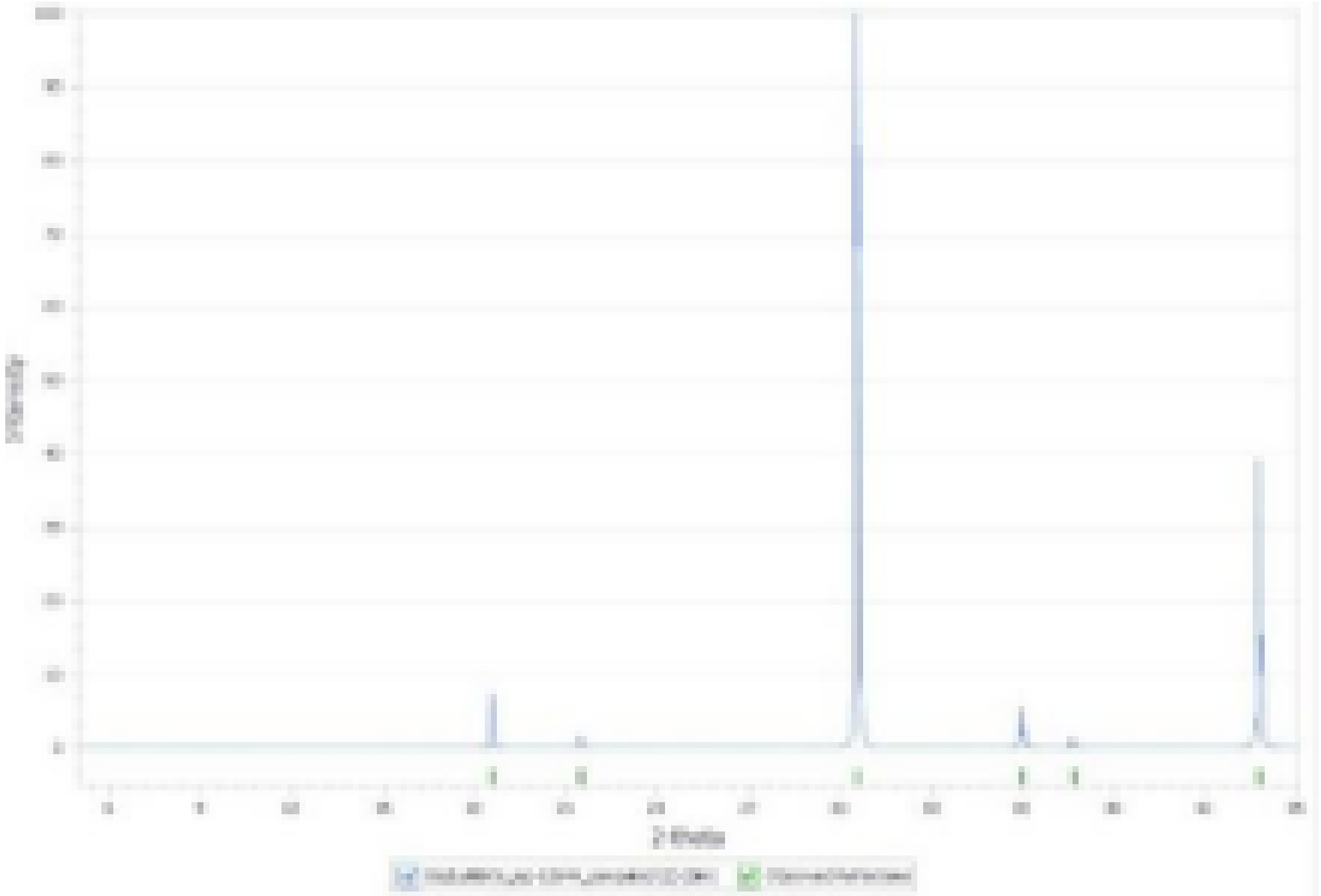

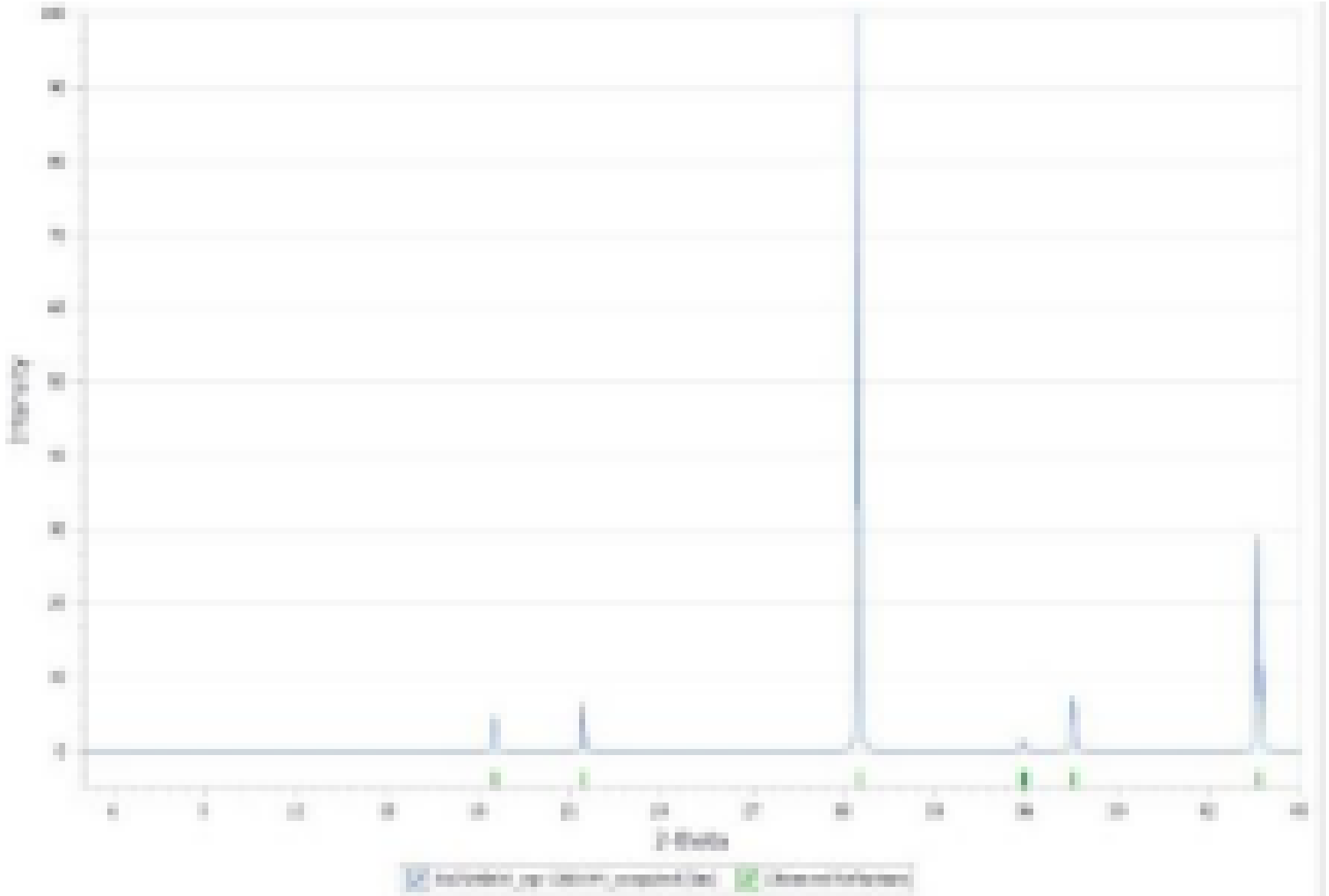